\documentclass{emulateapj}
\shorttitle{Repeating FRBs from Pulsar/Asteroid Impacts} \shortauthors{Dai, Wang, Wu \& Huang}

\begin{document}

\title{Repeating Fast Radio Bursts from Highly Magnetized Pulsars Travelling through Asteroid Belts}

\author{Z. G. Dai$^{1,2}$, J. S. Wang$^{1,2}$, X. F. Wu$^{3,4}$, \& Y. F. Huang$^{1,2}$}

\affil{$^1$School of Astronomy and Space Science, Nanjing University, Nanjing 210093, China; dzg@nju.edu.cn\\
$^2$Key Laboratory of Modern Astronomy and Astrophysics (Nanjing University), Ministry of Education, China\\
$^3$Purple Mountain Observatory, Chinese Academy of Sciences, Nanjing 210008, China\\
$^4$Joint Center for Particle, Nuclear Physics and Cosmology, Nanjing
University-Purple Mountain Observatory, Nanjing 210008, China}

\newcommand{\be}{\begin{equation}}
\newcommand{\ee}{\end{equation}}
\newcommand{\g}{\gamma}
\def\ba{\begin{eqnarray}}
\def\ea{\end{eqnarray}}
\def\cE{{\cal E}}
\def\cR{{\cal R}}
\def\bmu{{\mbox{\boldmath $\mu$}}}
\def\bphi{{\mbox{\boldmath $\phi$}}}

\begin{abstract}
Very recently \cite{Spitler2016} and \cite{Scholz2016} reported their detections of sixteen additional bright bursts
from the direction of the fast radio burst (FRB) 121102. This repeating FRB is inconsistent with all
the catastrophic event models put forward previously for hypothetically non-repeating FRBs. Here we
propose a different model, in which highly magnetized pulsars travel through asteroid belts of other stars.
We show that a repeating FRB could originate from such a pulsar encountering lots of asteroids
in the belt. During each pulsar-asteroid impact, an electric field induced outside the asteroid
has such a large component parallel to the stellar magnetic field that electrons are torn off the asteroidal
surface and accelerated to ultra-relativistic energies instantaneously. Subsequent movement of these electrons
along magnetic field lines will cause coherent curvature radiation, which can account for all the properties of an FRB.
In addition, this model can self-consistently explain the typical duration, luminosity, and repetitive rate
of the seventeen bursts of FRB 121102. The predicted occurrence rate of repeating FRB sources may imply that our model
would be testable in the next few years.
\end{abstract}


\section{Introduction}
Fast radio bursts (FRBs) are millisecond-duration flashes at typical frequency of order $\sim 1\,$GHz
\citep{Lorimer2007,Keane2012,Thornton2013,Burke2014,Spitler2014,Spitler2016,Ravi2015,Petroff2015,Champion2015,Masui2015,Keane2016}
The dispersion measures of total seventeen FRBs detected up to now are
in the range of a few hundreds to few thousands parsecs/cm$^3$, which strongly suggest that they are of
an extragalactic or even cosmological origin. Many models have been proposed to account for FRBs, including
giant flares from magnetars \citep{Popov2010,Kulkarni2014,Katz2016a}, giant pulses from
pulsars \citep{Connor2016,Cordes2016,Lyutikov2016}, eruptions of nearby flaring stars \citep{Loeb2014},
collisions between neutron stars and asteroids/comets \citep{Geng2015}, planetary companions around pulsars \citep{Mottez2014},
mergers of compact object binaries \citep{Totani2013,Kashiyama2013,Mingarelli2015,Wang2016,Liu2016,Zhang2016a,Zhang2016b,Punsly2016},
and collapses of supra-massive neutron stars to black holes \citep{Falcke2014,Zhang2014}.
A recent review of FRBs and relevant origin models can be seen in \cite{Katz2016b}.

Very recently \cite{Spitler2016} reported their detections of ten additional bright bursts
from FRB 121102 at 1.4 GHz at the Arecibo Observatory. On 17 May and 2 June 2015, two and eight bursts
were observed, respectively. \cite{Scholz2016} also reported their detections of the other bursts from this FRB, i.e.,
five with the Green Bank Telescope at 2 GHz and one at 1.4 GHz at the Arecibo Observatory. The latter six events
include one, four and one burst observed on 13 \& 19 November and 8 December 2015, respectively.
During the active periods, the rate of burst detections is $\sim 3\,{\rm h}^{-1}$ for bursts with flux density $>20\,$mJy.
Although these additional bursts and the original burst, which have consistent dispersion measures $\sim 556$\,pc/cm$^3$,
also strongly suggest an extragalactic origin, this FRB, due to its repetitiveness, is inconsistent with
all the catastrophic event models put forward previously for hypothetically non-repeating FRBs as listed above.

In this paper, we propose for the first time that a repeating FRB could originate from a
highly magnetized pulsar encountering lots of asteroids in an asteroid belt of another star
(for meanings and typical values of physical parameters and dependence of
observed quantities on some parameters in this model see Tables 1 and 2 respectively).
We analyze the impact physics and find that during each pulsar-asteroid impact a very strong electric field
is induced outside the asteroid when it is close to the stellar surface.
We show that because of this electric field, our model can well account for all the properties of FRB 121102.

Although the impact of a magnetized pulsar with an asteroid has been considered by \cite{Colgate1981}
for gamma-ray bursts (GRBs) and by \cite{Geng2015} for FRBs, the acceleration and radiation mechanisms
of ultra-relativistic electrons remain unknown. Our present work investigates such mechanisms and contains
several novel results. First, an electric field induced outside an elongated asteroid near the pulsar's
surface is found to have a large component parallel to the stellar magnetic field that accelerates electrons to
ultra-relativistic energies instantaneously, even though synchrotron radiation meanwhile cool them down.
Second, owing to subsequent curvature radiation, such electrons can emit
a radio signal. Third, we further find that coherent curvature radiation from these
electrons is consistent with cosmological FRBs. Fourth, we thus give order-of-magnitude estimates of both
the repetitive rate of an FRB and the occurrence rate of repeating sources in our model.
Finally, we show that the duration distribution and luminosity function of all the observed bursts of FRB 121102
can be well understood self-consistently within the framework of this model. To our knowledge, these results have not
been shown in the literature.

This paper is organized as follows.
In Section 2, we analyze the impact physics of an asteroid with a highly magnetized pulsar.
In Section 3, we discuss the properties of coherent curvature radiation.
In Section 4, we estimate the repetitive rate of an FRB and the occurrence rate
of repeating FRB sources. Finally, in Section 5, we present our conclusions and discussions.

\section{Impact Physics and Induced Electric Field}

\subsection{Duration and Luminosity of Energy Release}

The impact of an asteroid/comet with a neutron star was early proposed as an origin of GRBs
\citep{Harwitt1973,Colgate1981,VanBuren1981,Mitrofanov1990,Shull1995} and soft gamma-ray repeaters
\citep{Livio1987,Boer1989,Katz1994}. Following \cite{Colgate1981}, we assume that an asteroid
as a solid body falls freely in the gravitational field of a pulsar with mass $M$, radius $R_*$ and rotation
period $P_*$. This asteroid is originally approximated by a cylinder with radius $r_0$ and length $l_0=2r_0$.
Assuming that $s$ is the tensile strength of the asteroid and $\rho_0$ is its original mass density,
\cite{Colgate1981} found that the asteroid will be distorted tidally by the pulsar at the breakup radius,
\begin{eqnarray}
R_b & = & (\rho_0r_0^2GM/s)^{1/3}\nonumber \\ & = & 2.22\times 10^9m_{18}^{2/9}\rho_{0,0.9}^{1/9}s_{10}^{-1/3}(M/1.4M_\odot)^{1/3}\,{\rm cm},
\end{eqnarray}
where the convention $Q_x=Q/10^x$ in cgs units is adopted, $\rho_{0,0.9}=\rho_0/8\,{\rm g}\,{\rm cm}^{-3}=1$ and $s_{10}=1$ for
an iron-nickel asteroid with mass $m$, and $G$ is the gravitational constant.
The time difference of arrival of asteroidal leading and lagging fragments at the pulsar's surface is estimated
by\footnote{The factor $12/5$ in Equation (2) was erroneously written as $2/3$ by \cite{Colgate1981} and 2 by \cite{Geng2015}, respectively.}
\begin{eqnarray}
\Delta t & \simeq & \frac{12r_0}{5}\left(\frac{2GM}{R_b}\right)^{-1/2}\nonumber \\ & = & 1.58m_{18}^{4/9}\rho_{0,0.9}^{-5/18}
s_{10}^{-1/6}(M/1.4M_\odot)^{-1/3}\,{\rm ms}.
\end{eqnarray}
This timescale is not only independent of the stellar radius and weakly dependent on the other parameters,
but also it is consistent with the typical duration of an FRB. Furthermore, the average gravitational
energy release rate during $\Delta t$ is approximated by
\begin{eqnarray}
\dot{E}_{\rm G} \simeq \frac{GMm}{R\Delta t} & = & 1.18\times 10^{41}
m_{18}^{5/9}\rho_{0,0.9}^{5/18}s_{10}^{1/6}\nonumber \\ & & \times (M/1.4M_\odot)^{4/3}R_6^{-1}\,{\rm erg}\,{\rm s}^{-1}.
\end{eqnarray}
where $R$ is the distance of the asteroid to the pulsar's center. We see that this energy release rate is
consistent with the upper limit of the luminosity of a cosmological FRB for $m_{18}\sim\,{\rm a\,few}$,
$M\sim 1.4M_\odot$ and $R_6\sim 1$. These consistencies suggest that impacts of
asteroids with pulsars are a promising origin of FRBs.

In order to discuss the radiation properties of resultant FRBs in the next section,
we need to analyze how the asteroidal size will evolve with $R$.
The asteroid will be initially elongated as an incompressible flow from $R_b$ and subsequently further transversely compressed
at $R<R_i$. Here $R_i$ is the radius at which compression begins,
\begin{eqnarray}
R_i=\left(\frac{5s}{8P_0}\right)^{2/5}R_b & = & 2.91\times 10^8(\kappa/0.13)m_{18}^{2/9}
\rho_{0,0.9}^{1/9}\nonumber \\ & & \times s_{10}^{-1/3}(M/1.4M_\odot)^{1/3}\,{\rm cm},
\end{eqnarray}
where $P_0\simeq 100s$ is the solid body compressive strength and $\kappa\equiv(5s/8P_0)^{2/5}\simeq0.13$ \citep{Colgate1981}.
The asteroidal size has been found to evolve with $R$ through (1) length $l=l_0(R/R_b)^{-1/2}$ and radius $r=r_0(R/R_b)^{1/4}$
for $R_i\leq R\leq R_b$, and (2) $l=l_0(R/R_b)^{-1/2}$ and $r=r_0(R_i/R_b)^{1/4}(R/R_i)^{1/2}$ for $R\leq R_i$
\citep{Colgate1981}. The latter evolution thus gives the length and radius of the asteroid approaching the pulsar's surface,
\begin{eqnarray}
l & = & 2.55\times 10^7m_{18}^{4/9}\rho_{0,0.9}^{-5/18}s_{10}^{-1/6}\nonumber \\ & & \times (M/1.4M_\odot)^{1/6}R_6^{-1/2}\,{\rm cm},
\end{eqnarray}
\begin{eqnarray}
r & = & 9.55\times 10^3(\kappa/0.13)^{-1/4}m_{18}^{2/9}
\rho_{0,0.9}^{-7/18}\nonumber \\ & & \times s_{10}^{1/6}(M/1.4M_\odot)^{-1/6}R_6^{1/2}\,{\rm cm},
\end{eqnarray}
and the asteroidal mass density
\begin{eqnarray}
\rho\equiv\frac{m}{\pi r^2l} & = & 137(\kappa/0.13)^{1/2}m_{18}^{1/9}\rho_{0,0.9}^{19/18}
s_{10}^{-1/6}\nonumber \\ & & \times (M/1.4M_\odot)^{1/6}R_6^{-1/2}\,{\rm g}\,{\rm cm}^{-3}.
\end{eqnarray}

As the asteroid falls freely, its movement will be eventually affected by the magnetic field of the pulsar.
We define the Alfv\'en radius at which the asteroidal kinetic energy density ($\rho v_{\rm ff}^2/2$)
is equal to the magnetic energy density of the pulsar ($\mu^2/8\pi R^6$),
\begin{eqnarray}
R_A & = & 1.10\times 10^6(\kappa/0.13)^{-1/9}m_{18}^{-2/81}\rho_{0,0.9}^{-19/81}\nonumber \\ & & \times
s_{10}^{1/27}(M/1.4M_\odot)^{-7/27}\mu_{30}^{4/9}\,{\rm cm},
\end{eqnarray}
where $v_{\rm ff}=(2GM/R)^{1/2}$ is the free-fall velocity and $\mu$ is the magnetic dipole moment of the pulsar.
We have used Equation (7) to derive Equation (8). The inequality $R_A\lesssim R_*$ is required in order that
an effect of the magnetic field on the asteroidal falling is insignificant and that our next derivations are self-consistent.
This inequality is valid only under the assumption that
\begin{eqnarray}
\mu_{30} & \lesssim & 0.807(\kappa/0.13)^{1/4}m_{18}^{1/18}\rho_{0,0.9}^{19/36}\nonumber \\ & & \times s_{10}^{-1/12}(M/1.4M_\odot)^{7/12}R_{*,6}^{9/4}.
\end{eqnarray}
Of course, an actual upper limit of $\mu_{30}$ might be looser than this assumption.

\begin{figure}
\begin{center}
\vspace{-15mm}
\includegraphics[scale=0.36]{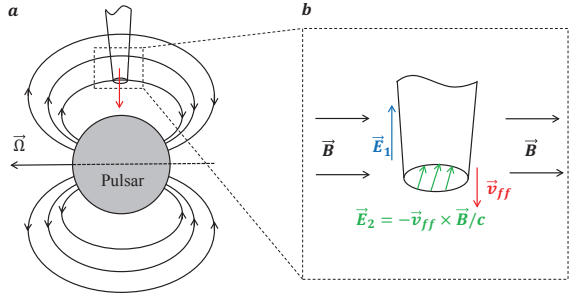}
\caption{Schematic picture of the impact of a highly magnetized pulsar with a radially elongated, transversely compressed,
freely falling asteroid. {\bf Panel a:} The asteroid penetrates through the magnetosphere of the pulsar that rotates with
angular velocity of ${\bf \Omega}$. {\bf Panel b:} Two components of an induced electric field are generated. As the pulsar spins,
the asteroid crosses the magnetic field lines transversely, giving rise to the first component,
${\bf E}_1=-{\bf \Omega}\times{\bf R}\times{\bf B}/c$, whose direction is shown by the blue arrow.
As it falls freely, the asteroid crosses the magnetic field lines longitudinally, leading to the second component,
${\bf E}_2=-{\bf v}_{\rm ff}\times{\bf B}/c$, whose direction is shown by the green arrows. When the asteroid is close to
the pulsar's surface, $|{\bf E}_1|\ll |{\bf E}_2|$ and the induced electric field outside the asteroid has such a large component
parallel to the magnetic field that electrons will be torn off the asteroidal surface and accelerated to ultra-relativistic energies,
as shown in the text.}
\end{center}
\end{figure}

\subsection{Induced Electric Field}

The radially elongated, transversely compressed asteroid penetrates through the stellar magnetosphere,
resulting in two components of an induced electric field. First, as the pulsar spins with angular velocity
$\Omega=2\pi/P_*$, the asteroid crosses the magnetic field lines transversely. This gives rise to
one component of the induced electric field, ${\bf E}_1=-{\bf \Omega}\times{\bf R}\times{\bf B}/c$ (see Fig. 1).
The strength of this component is
\be
E_1=2.10\times 10^8(P_*/1\,{\rm s})^{-1}\mu_{30}R_6^{-2}\,{\rm volt}\,{\rm cm}^{-1}.
\ee
Second, as it falls freely, the asteroid crosses the magnetic field lines longitudinally. This leads to the other
component of the induced electric field, ${\bf E}_2=-{\bf v}_{\rm ff}\times{\bf B}/c$ (see Fig. 1), whose strength reads
\be
E_2=6.44\times 10^{11}(M/1.4M_\odot)^{1/2}\mu_{30}R_6^{-7/2}\,{\rm volt}\,{\rm cm}^{-1}.
\ee
Comparing Equations (10) and (11), we see that $E_1\ll E_2$ for $R_6\sim 1$, even if $P_*$ is as short as 1\,ms.

If the pulsar is highly magnetized, rapid separations of electrons and ions in the interior of the freely-falling asteroid,
owing to opposing Lorentz forces produced when the asteroid moves across the pulsar's
magnetic field, lead to the instantaneous appearance of net charge near the asteroidal surface. The same charge-layering
process is well known in the case of a bounded plasma moving across a magnetic field at a sub-Alfv\'enic velocity
\citep{Chandra1960,Schmidt1960}. The result of this process is that while the induced electric field (which forms
the so-called Hall voltage) within the asteroid guarantees force-free ${\bf E}_2+{\bf v}_{\rm ff}\times{\bf B}/c=0$,
the stray electric field outside the asteroid must have a large component parallel to the magnetic field so that
electrons are torn off the asteroidal surface and accelerated to relativistic energies along the magnetic field lines.
This physical phenomenon has been observed in numerical simulations by \cite{Galvez1991} and \cite{Neubert1992},
who found that charge layers can indeed accelerate electrons to relativistic energies for plasma flows moving
at a sub-Alfv\'enic velocity. Of course, protons near the stellar surface would also be accelerated
simultaneously to high-energy cosmic rays \citep{Litwin2001}, which could further produce high-energy neutrinos by
photo-meson interactions.

The movement of relativistic electrons along the magnetic field lines causes an electric circuit connecting the pulsar and
asteroid. What we should point out is that a unipolar induction DC circuit model was originally put forward in the Jupiter-Io
system \citep{GoldreichL1969}. This model was then successfully applied to several astrophysical systems, such as white dwarf-white dwarf
binaries \citep{Wu2002} and neutron star-neutron star/black hole binaries \citep{Hansen2001,McWilliams2011,Piro2012,Lai2012}.
Recently, \cite{Wang2016} argued the final inspiral of two neutron stars as an origin of non-repeating FRBs within the framework
of this model and suggested interesting associations of FRBs with short GRBs and gravitational wave events. Therefore,
the merger of two neutron stars might give birth to ``triplets'', among which the eldest is a gravitational wave burst, the youngest
is a short GRB, and the intermediate is an FRB.

\section{Radiation Properties}

In this section, we first investigate acceleration of electrons outside the elongated asteroid
near the stellar surface. Owing to the electric field parallel to the magnetic field that is of the order $E_2$, charged particles
will be torn off the asteroidal surface and create a new plasma near the asteroid. As discussed in Subsection 2.2, electrons
in this plasma will be immediately accelerated to ultra-relativistic energies by the electric field, but meanwhile they
will also be cooled down by synchrotron radiation in the strong magnetic field of the pulsar. The maximum Lorentz factor of
an accelerated electron is determined by assuming that its synchrotron radiation power ($\sigma_Tc\gamma_{\rm max}^2B^2/6\pi$)
is equal to the power due to electric field acceleration ($eE_2c$), that is,
\be
\gamma_{\rm max}=\left(\frac{6\pi e E_2}{\sigma_TB^2}\right)^{1/2}=93.6(M/1.4M_\odot)^{1/4}\mu_{30}^{-1/2}R_6^{5/4},
\ee
where $\sigma_T$ is the Thomson scattering cross section. The acceleration timescale is given by
$\tau_{\rm acc}=\gamma_{\rm max}m_ec/eE_2=8.25\times 10^{-18}(M/1.4M_\odot)^{-1/4}\mu_{30}^{-3/2}R_6^{19/4}\,{\rm s}$,
whose corresponding length ($\simeq c\tau_{\rm acc}$) is much shorter than the transverse size of the acceleration region ($\simeq 2r$).
This implies that the induced electric field can indeed accelerate electrons to $\gamma_{\rm max}$ instantaneously.

According to Poisson's equation, the electric field ${\bf E}_2$ in the acceleration region
outside and near the asteroid is associated with the charge density \citep{GoldreichJ1969,Shapiro1983},
$\rho_e=(1/4\pi)\nabla\cdot {\bf E}_2$, which is similar to the well-known Goldreich-Julian density in the magnetosphere
of a pulsar and thus gives the average charge number density,
\begin{eqnarray}
n_e & = & \frac{1}{4\pi e}\nabla\cdot {\bf E}_2\simeq\frac{7\mu r\sqrt{2GM}}{8\pi ec R^{11/2}}\nonumber \\
& = & 3.57\times 10^{12}(\kappa/0.13)^{-1/4}m_{18}^{2/9}\rho_{0,0.9}^{-7/18}
\nonumber \\ & & \times s_{10}^{1/6}(M/1.4M_\odot)^{1/3}\mu_{30}R_6^{-5}\,{\rm cm}^{-3},
\end{eqnarray}
where $r$ and $E_2$ have been substituted by Equations (6) and (11) respectively. In addition to the electric circuit connecting the asteroid
and pulsar, the movement of ultra-relativistic electrons along the magnetic field lines will result in coherent curvature radiation.

We next discuss the radiation properties. If it moves along a magnetic field line with curvature
radius $\rho_c$, an ultra-relativistic electron with Lorentz factor of $\gamma$ will produce curvature radiation,
whose characteristic frequency is given by
\be
\nu_{\rm curv}={3c\gamma^3\over 4\pi \rho_c}=7.16\times 10^3\gamma^3 \rho_{c,6}^{-1}\,\rm{Hz}.
\ee
For a typical FRB with $\nu_{\rm curv}$ of order $\sim 1$\,GHz, it is required that $\gamma\sim 52\rho_{c,6}^{1/3}
(\nu_{\rm curv}/1\,{\rm GHz})^{1/3}$, which is basically in agreement with Equation (12).
The power of curvature radiation from an ultra-relativistic electron is $P_c=2\g^4e^2c/3\rho_c^2$.
Theoretically, if the radiation wavelength is comparable to the size of the emitting region,
the curvature radiation becomes coherent \citep{Ruderman1975}. Observationally,
cosmological FRBs have extremely high brightness temperatures $\sim 10^{37}\,$K, requiring coherent radiation
by ``bunches'' of electrons \citep{Katz2014,Cordes2016}.

Considering an emitting region enclosed by the electric circuit and all emitting electrons
having bulk movement along the magnetic field lines, the total luminosity of coherent radiation is
$L_{\rm tot}\simeq (P_c N_{\rm coh}^2)\times N_{\rm slice}$, where $N_{\rm coh}$ is the number of bunching
electrons in a slice producing coherent radiation, and $N_{\rm slice}=V_{\rm tot}/V_{\rm slice}$ is
the number of coherent slices. Here $V_{\rm tot}$ and $V_{\rm slice}$ are
the total volume of the emitting region and the volume of a slice, respectively. In our model, $V_{\rm tot}\sim
\pi R\times 2r\times \zeta R$, where $\zeta R$ is assumed to be the height of the emitting region
with factor $\zeta\sim [\pi R/(2c)]\times(v_{\rm ff}/R) \sim 1.0(M/1.4M_\odot)^{1/2}R_6^{-1/2}$,
and $V_{\rm slice}\sim 2r\times \zeta R\times (c/\nu_{\rm curv})$, so we have $N_{\rm slice}\sim 3\gamma^3R/(4\rho_c)$,
which is similar to the estimate of \cite{Falcke2014}. Since the total number of emitting electrons
is approximated by $N_{e,{\rm tot}}\simeq N_{\rm coh} N_{\rm slice}\sim V_{\rm tot}n_e$, we derive the total luminosity
of coherent curvature radiation,
\begin{eqnarray}
L_{\rm tot} & \simeq & P_c N_{e,{\rm tot}}^2N_{\rm slice}^{-1}\nonumber \\
& \sim & 2.63\times 10^{40}(\kappa/0.13)^{-1}m_{18}^{8/9}\rho_{0,0.9}^{-14/9}s_{10}^{2/3}\nonumber \\ & & \times
(M/1.4M_\odot)^{19/12}\mu_{30}^{3/2}R_6^{-23/4}\rho_{c,6}^{-1}\,{\rm erg}\,{\rm s}^{-1},
\end{eqnarray}
where $\gamma$ has been substituted by Equation (12). A strong dependence of $L_{\rm tot}$ on $R$
(i.e., $L_{\rm tot}\propto R^{-23/4}$) implies that a significant fraction of the pulsar-asteroid gravitational binding
energy will be converted to a radio signal via curvature radiation only when the asteroid approaches
the surface of the pulsar. Therefore, a required consistency of Equation (15) with the lower limit of an observed FRB
luminosity ($L_{\rm min}$) implies the assumption that
\begin{eqnarray}
\mu_{30} & \gtrsim & 0.113(\kappa/0.13)^{2/3}m_{18}^{-16/27}\rho_{0,0.9}^{28/27}s_{10}^{-4/9}\nonumber \\ & & \times (M/1.4M_\odot)^{-19/18}
R_{*,6}^{23/6}\rho_{c,6}^{2/3}L_{\rm min,39}^{2/3},
\end{eqnarray}
where $L_{\rm min,39}=L_{\rm min}/10^{39}{\rm erg}\,{\rm s}^{-1}$. Under this assumption, Equations (15) together with
Equation (2) can well account for the luminosity and duration of an FRB.

Furthermore, the radially elongated asteroid, during its falling, could further bend the magnetic field lines, leading to a smaller curvature
radius $\rho_c$ and a larger luminosity $L_{\rm tot}$. However, from the point of view of energy conservation, the coherent curvature
radiation discussed above is powered by the pulsar-asteroid gravitational binding energy, and thus the upper
limit of the emission luminosity can be approximately given by Equation (3). We see that this power is enough available for an FRB
if typical values of the model parameters are taken in Table 1.

\section{Order-of-Magnitude Rate Estimates}

We now estimate the repetitive rate of an FRB. Under the assumption that an originally spherical asteroid with radius
$r_a=(3/2)^{1/3}r_0$ and average proper velocity $v_a$ impacts with a pulsar with average proper velocity $v_*$,
their impact cross section can be described by the famous formula of \cite{Safronov1972},
\be
\sigma_a=\pi(R_*+r_a)^2(1+2\theta),
\ee
where $\theta\equiv G(M+m)/[(R_*+r_a)v_{\rm rel}^2]$ is the Safronov number and
$v_{\rm rel}=|{\bf v}_*-{\bf v}_a|$ is the velocity of the asteroid relative to the pulsar. For $m\ll M$,
$r_a\sim R_*$, $v_a\ll v_*$ \citep[for a review see][]{Davis2002}, and
$v_{\rm rel}\simeq v_*$, we have
\be
\sigma_a\sim \frac{4\pi GMR_*}{v_*^2}\sim 2.35\times 10^{19}R_{*,6}(M/1.4M_\odot)v_{*,7}^{-2}\,{\rm cm}^2.
\ee
By further assuming that the inner radius of an asteroid belt is $R_a$, the width and thickness of the belt are both approximated
by $\eta R_a$ with $\eta\ll 1$, and the total number of asteroids in the belt is $N_a$, we estimate the
impact rate as
\begin{eqnarray}
{\cal{R}}_a & \sim & \frac{\sigma_a v_*N_a}{2\pi \eta^2 R_a^3} \sim 1.25R_{*,6}
(M/1.4M_\odot)v_{*,7}^{-1}\nonumber \\ & & \times N_{a,10}(\eta/0.2)^{-2}(R_a/2{\rm AU})^{-3}\,{\rm h}^{-1}.
\end{eqnarray}
This rate is well consistent with the observed burst rate of FRB 121102 ($\sim 3\,{\rm h}^{-1}$), if we take
$v_{*,7}\sim 2$ \citep{Blaes1993,Ofek2009}, $\eta\sim 0.2$, $R_a\sim 2\,{\rm AU}$, and $N_{a,10}\sim 5$.
The latter three parameters are reasonable for both solar and extrasolar asteroid belts, and in particular, the last parameter
means a heavy belt mass $\sim 8(N_{a,10}/5)m_{18}M_\oplus$, indicating an asteroid belt younger than $10^8$\,years
\citep{Jones2007,Zhou2016}.

We next estimate the occurrence rate of repeating FRB sources. Assuming $N_{\rm b}$ and $N_*$ to be the total numbers
of asteroid belts and pulsars in a galaxy with typical effective volume $V_{\rm gal}$ respectively,
we consider face-on collisions between asteroid belts and pulsars and find the rate of such collisions
\begin{eqnarray}
{\cal{R}}_{b,{\rm face}} & \sim & 2\pi \eta R_a^2v_*(N_{\rm b}/V_{\rm gal})N_*\nonumber \\
& \sim & 1.20\times 10^{-7}(\eta/0.2)(R_a/2{\rm AU})^2
v_{*,7}\nonumber \\ & & \times N_{{\rm b},11}N_{*,8}V_{\rm gal,12}^{-1}\,{\rm galaxy}^{-1}\,{\rm yr}^{-1},
\end{eqnarray}
where $V_{\rm gal,12}=V_{\rm gal}/10^{12}\,{\rm pc}^3$ and every star is assumed
to host only one asteroid belt and all the stars are uniformly distributed in the galaxy.
This rate and the edge-on collision rate estimated below are clearly conservative values (i.e., lower limits). One reason
for this argument is that stars mainly concentrate on star-forming regions in the galaxy and thus the number density
of asteroid belts in these regions should be larger than the one assumed above. The other reason is that one star may
host a few asteroid belts, as in the solar system \citep[for a review see][]{DeMeo2014}. For one face-on collision, there is
only one travel through the belt. This gives the repetitive number of an FRB, $n_{\rm burst}\sim \sigma_aN_a/(2\pi\eta R_a^2)
\sim 210R_{*,6}(M/1.4M_\odot)v_{\rm rel,7}^{-2}N_{a,10}(R_a/2{\rm AU})^{-2}$. Furthermore,
we calculate the rate of edge-on collisions between asteroid belts and pulsars in a galaxy,
\begin{eqnarray}
{\cal{R}}_{b,{\rm edge}} & \sim & 2\eta R_a^2v_*(N_{\rm b}/V_{\rm gal})N_*\nonumber \\
& \sim & 3.82\times 10^{-8}(\eta/0.2)(R_a/2{\rm AU})^2
v_{*,7}\nonumber \\ & & \times N_{{\rm b},11}N_{*,8}V_{\rm gal,12}^{-1}\,{\rm galaxy}^{-1}\,{\rm yr}^{-1}.
\end{eqnarray}
The total rate ${\cal{R}}_b\equiv {\cal{R}}_{b,{\rm face}}+{\cal{R}}_{b,{\rm edge}}\sim 2\times 10^{-6}\,$per galaxy per year,
suggesting that at least $\sim 2\times 10^3$ repeating FRB sources per year would be detectable,
provided that $N_{*,8}\sim 2.5$ and $v_{*,7}\sim 2$ \citep{Blaes1993,Ofek2009},
$N_{{\rm b},11}\sim 2$ and $V_{\rm gal,12}\sim 0.95$ \citep{Shull1995}, and that the cosmological
comoving volume at redshift $z\lesssim 1$ contains $N_{\rm gal}\sim 10^9$ late-type galaxies \citep{Madgwick2002,Thornton2013}.
In addition, one edge-on collision implies two travels through the belt, which result in two clusters of bursts. The time interval
between them is $\sim R_a/v_*\sim 17(R_a/2{\rm AU})(v_{*,7}/2)^{-1}\,$days, which is consistent with
the observed one \citep[$\sim 15\,$days,][]{Spitler2016}.

The radius of a pulsar capturing a host star at the center of an asteroid belt is
\begin{eqnarray}
R_{\rm capture}& \sim & 2G(M+M_{\rm star})/v_*^2\nonumber \\ & \sim & 1.6\times 10^{12}\left(\frac{M+M_{\rm star}}{2.4M_\odot}\right)(v_{*,7}/2)^{-2}\,{\rm cm},
\end{eqnarray}
where $M_{\rm star}$ is the mass of the star. If the impact parameter of this pulsar with the star is
less than $R_{\rm capture}$, they will form a self-gravitational binding (binary) system. In such a case, this pulsar could
collide with the asteroid belt {\em back and forth}, so that many clusters of repeating bursts would be expected to occur
in the case of edge-on collisions. This feature is an observational, unique signature for our model.

\section{Conclusions and Discussions}

\cite{Spitler2016} and \cite{Scholz2016} have found that FRB 121102, due to its repeating nature,
is inconsistent with all the catastrophic event models listed in Section 1. In this paper, we have proposed a novel
model, in which a repeating FRB could originate from a highly magnetized pulsar encountering lots of
asteroids in an asteroid belt of another star (for a summary of the model parameters and observed quantities
see Tables 1 and 2). During each pulsar-asteroid impact, an electric field induced outside the asteroid
has a very large component parallel to the stellar magnetic field that accelerates electrons to ultra-relativistic
energies instantaneously. We show that subsequent movement of these electrons along the magnetic field lines will
cause coherent curvature radiation, which can well account for the emission frequency, duration, and luminosity
of a cosmological FRB.

Furthermore, the repetitive rate estimated based on our model is high enough to explain the repetitiveness
of FRB 121102. The theoretical time interval among a few clusters of bursts for edge-on collisions is
consistent with the observed one. The predicted rate of encounters between pulsars and asteroid belts
indicates that at least $\sim 2\times 10^3$ repeating FRB sources per year would be detectable. This may imply that our model
would be testable if a new repeating FRB is discovered in the next few years. In particular, if a pulsar captures a star
successfully and collides with an asteroid belt around this star back and forth, many clusters of repeating bursts
would be expected to occur in the case of edge-on collisions. This feature, if observed, will be possible evidence for our model.

Fortunately, this feature has been probably shown by the observations on FRB 121102. Let's now discuss
in a little detail how our model explains total seventeen bursts of this FRB: a highly magnetized
pulsar with proper velocity of $v_*$ could first edge-on encounter an outer Kuiper-like belt around a main-sequence star.
If the outer radius of this Kuiper-like belt is $R_K\sim 100\,$AU as in the solar system \citep{Jones2007,Zhou2016}, then
a typical traveling timescale of the pulsar in this belt is $\sim R_K/v_*=2.4(R_K/100\,{\rm AU})(v_{*,7}/2)^{-1}\,$yr,
in which the pulsar could be impacted by a massive asteroid, leading to the original burst of FRB 121102. After this timescale,
the pulsar could edge-on encounter an inner asteroid belt (with an inner radius of $R_a$) around the star
as also in the solar system \citep{DeMeo2014}. If the impact parameter of the pulsar with the star
is less than the capture radius $R_{\rm capture}$, then the two stars could form a binary system and move around
their center of mass on a probably highly elliptical orbit, in which case the pulsar could always collide with
the asteroid belt back and forth. The time interval in which the pulsar edge-on encounters the asteroid belt twice
is $\sim R_a/v_*=17(R_a/2\,{\rm AU})(v_{*,7}/2)^{-1}\,$days, which is consistent with the one observed by
\cite{Spitler2016} and \cite{Scholz2016}. This result supports our model. In addition, the detectable rate of
FRB 121102-like sources is approximated by ${\cal{R}}_{121102}\sim (R_{\rm capture}^2/\eta R_a^2)
{\cal{R}}_bN_{\rm gal}\sim 30[(M+M_{\rm star})/2.4M_\odot]^2\,{\rm yr}^{-1}$,
which is strongly dependent on the total binary mass if typical values of the other parameters are taken in Table 1.

We next discuss a recent statistic analysis of \cite{WangYu2016}, who found
the cumulative duration ($T\equiv\Delta t$) distribution of all the observed bursts of FRB 121102,
$N(>T)\propto T^{-\alpha_T}$ with $\alpha_T=0.95\pm 0.32$. This distribution, together with Equation (2),
requires that the differential size distribution ($dN/dD\propto D^{-\beta}$) of asteroids with diameter $D\lesssim 10$\,km
has an index $\beta=(4\alpha_T+3)/3= 2.26\pm 0.43$, which is well consistent with both the Sloan Digital Sky
Survey (SDSS) data \citep[$\beta\simeq2.30$,][]{Ivezic2001,Davis2002} and the Subaru Main Belt Asteroid Survey (SMBAS) data
\citep[$\beta\simeq2.29$,][]{Yoshida2007} of solar system objects. \cite{WangYu2016} also gave the luminosity function of these bursts,
$N(>L)\propto L^{-\alpha_L}$ with index of $\alpha_L=0.78\pm 0.16$. Assuming that $R\simeq R_*$ and $\rho_c={\rm constant}$
in Equation (15) for all the bursts, this index is theoretically related with $\beta$ through $\beta=(8\alpha_L+3)/3=3.08\pm 0.43$,
which is consistent with the index of the differential size distribution of asteroids derived from $\alpha_T$ and
marginally consistent with the SDSS and SMBAS data of solar system objects. These results not only further support our model,
but also show its self-consistency.

We further discuss the other possible phenomenon in the pulsar-asteroid encounter model.
When an asteroid impacts the surface of a pulsar, a resultant hot spot with radius given approximately by Equation (6)
is powered by the gravitational energy release, but meanwhile it is cooled down by the surface black-body radiation. Under the assumption of
thermal equilibrium, the temperature of the hot spot is estimated by $T_{\rm spot}\sim (\dot{E}_{\rm G}/\sigma_{\rm SB}\pi r^2)^{1/4}
=1.64\times 10^9m_{18}^{1/36}\rho_{0,0.9}^{19/72}s_{10}^{-1/24}(M/1.4M_\odot)^{5/12}R_{*,6}^{-1/2}\,$K, where $\sigma_{\rm SB}$
is the Stefan-Boltzmann constant. We see that this black-body temperature is very weakly dependent on asteroidal mass and
that the emission from the hot spot turns out to be at soft gamma-ray energy rather than X-ray energy suggested by \cite{Huang2014}
and \cite{Geng2015}. This is the reason that pulsar-asteroid impacts were widely argued to be an origin of nearby GRBs
in the early literature. Assuming isotropic emission, the farthest distance of such GRBs is estimated by $D_{\rm GRB,max}\sim
(\dot{E}_{\rm G}/4\pi F_\gamma)^{1/2}=31.4m_{18}^{5/18}\rho_{0,0.9}^{5/36}s_{10}^{1/12}(M/1.4M_\odot)^{2/3}R_{*,6}^{-1/2}
F_{\gamma,-12}^{-1/2}\,$Mpc, where $F_\gamma=F_{\gamma,-12}\times 10^{-12}\,{\rm erg}\,{\rm cm}^{-2}\,{\rm s}^{-1}$ is the sensitivity
of a detector. Further assuming the total rate of FRBs ${\cal R}_{\rm FRB}\sim 10^4\,{\rm sky}^{-1}\,{\rm day}^{-1}$
and their farthest distance $D_{\rm FRB,max}\sim 3.2$\,Gpc \citep{Thornton2013},
we obtain the observable rate of GRBs, ${\cal R}_{\rm GRB}\sim {\cal R}_{\rm FRB}
(D_{\rm GRB,max}/D_{\rm FRB,max})^3\sim 3.44m_{18}^{5/6}\rho_{0,0.9}^{5/12}s_{10}^{1/4}(M/1.4M_\odot)^2R_{*,6}^{-3/2}
F_{\gamma,-12}^{-3/2}\,{\rm yr}^{-1}$, if all FRBs result from pulsar-asteroid impacts. This implies
that only a few extremely-low-luminosity GRBs associated with FRBs per year would be detected by a satellite with sensitivity of
$F_\gamma=10^{-12}\,{\rm erg}\,{\rm cm}^{-2}\,{\rm s}^{-1}$.

Finally, what we want to point out is that quasi-periodic FRB repetitions should not occur in our model.
The reason for this argument is that asteroids not only show a highly nonuniform distribution within their belt but also usually have
a steep size distribution. Therefore, encounters of a pulsar and asteroids with $m_{18}\sim\,$ a few are not quasi-periodic.

\acknowledgements
We thank an anonymous referee for valuable comments and constructive suggestions that have allowed us to improve our manuscript.
We also thank Ji-Lin Zhou and Li-Yong Zhou for helpful suggestions on asteroid belts, and Anthony L. Piro, Zhi-Qiang Shen,
Xiang-Yu Wang and Bing Zhang for useful discussions on FRBs. This work was supported by the National Basic Research Program
(``973" Program) of China (grant No. 2014CB845800) and the National Natural Science Foundation of China (grant Nos. 11573014,
11322328 and 11473012). X.F.W. was also partially supported by the Youth Innovation Promotion Association (No. 2011231) and
the Strategic Priority Research Program ``The Emergence of Cosmological Structure'' (grant No. XDB09000000) of the Chinese
Academy of Sciences.

\clearpage
\begin{table}
\caption{Meanings and typical values of physical parameters in the pulsar-asteroid-belt encounter model.}
\begin{center}
\begin{tabular}{lcccc}
\hline\hline
Meaning & Symbol & Typical value & References \\ \hline
{\em A pulsar} \\ \hline
Mass & $M$ & $1.4M_\odot$ & [1] \\
Radius & $R_*$ & $10^6$\,cm & [1] \\
Rotation period & $P_*$ & $\lesssim {\rm a\,few}\,{\rm seconds}$ & [1] \\
Magnetic dipole moment$^\dag$ & $\mu$ & ${\rm a\,few\,} 10^{29}\,{\rm G}\,{\rm cm}^3$ & [2] \\
Average proper velocity & $v_*$ & $\sim 2\times 10^7\,{\rm cm}\,{\rm s}^{-1}$ & [3,4] \\
Number of pulsars & $N_*$ & $\sim 2.5\times 10^8/{\rm galaxy}$ & [3,4] \\ \hline
{\em An iron-nickel asteroid} \\ \hline
Mass$^\ddag$ & $m$ & ${\rm a\,few}\,10^{18}$\,g & [2,5] \\
Original density & $\rho_0$ & $8\,{\rm g}\,{\rm cm}^{-3}$ & [5] \\
Original cylindrical radius & $r_0$ & $2.71m_{18}^{1/3}\rho_{0,0.9}^{-1/3}\,$km & [5] \\
Original cylindrical length & $l_0$ & $5.42m_{18}^{1/3}\rho_{0,0.9}^{-1/3}\,$km & [5] \\
Tensile strength & $s$ & $\sim 10^{10}\,{\rm dyn}\,{\rm cm}^{-2}$ & [5] \\
Compressive strength & $P_0$ & $\sim 10^{12}\,{\rm dyn}\,{\rm cm}^{-2}$ & [5] \\
Factor $\kappa\equiv(5s/8P_0)^{2/5}$ & $\kappa$ & $\sim 0.13$ & [5]  \\
Average proper velocity & $v_a$ & $\ll 2\times 10^7\,{\rm cm}\,{\rm s}^{-1}$ & [6] \\ \hline
{\em Curvature radiation} \\ \hline
Curvature radius near pulsar$^\S$ & $\rho_c$ & $\sim 10^6\,$cm & [7] \\
Height factor of emitting region & $\zeta$ & $\sim 1.0R_6^{-1/2}\,$ & [2] \\ \hline
{\em An asteroid belt} & & \\ \hline
Inner radius & $R_a$ & $\sim 2$\,AU & [8] \\
Width or thickness factor & $\eta$ & $\sim 0.2$ & [8] \\
Mass of a host star & $M_{\rm star}$ & $\lesssim {\rm few\,tens\,}M_\odot$ & [2,9] \\
Number of asteroids & $N_a$ & $\sim 5\times 10^{10}/{\rm belt}$ & [10,11] \\
Number of asteroid belts & $N_b$ & $\sim 2\times10^{11}/{\rm galaxy}$ & [9] \\ \hline
{\em A galaxy} \\ \hline
Effective volume & $V_{\rm gal}$ & $\sim 0.95\times 10^{12}\,{\rm pc}^3$ & [9] \\
Number of galaxies & $N_{\rm gal}$ & $\sim 10^9$ at $z\lesssim 1$ & [12,13] \\ \hline\hline
\end{tabular}
\end{center}
Notes: $^\dag$The upper and lower limits of the magnetic dipole moment of a pulsar are given by Equations (9) and (16),
respectively; $^\ddag$this typical mass of an asteroid is required to account for the duration and luminosity of
a cosmological FRB, but less massive asteroids only produce faint FRBs while much more massive
asteroids have a too low population to explain the observed rate of FRBs; $^\S$a smaller curvature radius, possibly
resulting from asteroidal falling, leads to a larger emission luminosity, as discussed in Section 3. References:
[1] \cite{Shapiro1983}, [2] this paper, [3] \cite{Blaes1993}, [4] \cite{Ofek2009}, [5] \cite{Colgate1981},
[6] \cite{Davis2002}, [7] \cite{Ruderman1975}, [8] \cite{DeMeo2014}, [9] \cite{Shull1995}, [10] \cite{Jones2007},
[11] \cite{Zhou2016}, [12] \cite{Madgwick2002}, and [13] \cite{Thornton2013}.
\end{table}

\begin{table}
\caption{Dependence of observed quantities on some parameters in the pulsar-asteroid-belt encounter model.}
\begin{center}
\begin{tabular}{lccc}
\hline\hline
Description & Symbol & Dependence \\ \hline
Duration of an FRB$^\sharp$ & $\Delta t$ & $1.58m_{18}^{4/9}\,$ms  \\
Luminosity of an FRB$^\sharp$ & $L_{\rm tot}$ & $\sim 2.63\times 10^{40}m_{18}^{8/9}\,{\rm erg}\,{\rm s}^{-1}$  \\
Total energy release rate$^\sharp$ & $\dot{E}_{\rm G}$ & $\sim 1.18\times 10^{41}m_{18}^{5/9}\,{\rm erg}\,{\rm s}^{-1}$ \\
Duration distribution$^\P$ & $N(>T)$ & $\propto T^{-\alpha_T}$, where $\alpha_T=3(\beta-1)/4$ \\
Luminosity function$^\P$  & $N(>L)$ & $\propto L^{-\alpha_L}$, where $\alpha_L=3(\beta-1)/8$ \\
Repetitive rate$^\|$ & ${\cal{R}}_a$ & $\sim3.12(v_{*,7}/2)^{-1}(N_{a,10}/5)\,{\rm h}^{-1}$  \\
Occurrence rate$^\|$ & ${\cal{R}}_b$ & $\sim2.0\times 10^{-6}(v_{*,7}/2)(N_{{\rm b},11}/2)(N_{*,8}/2.5)\,{\rm gal}^{-1}\,{\rm yr}^{-1}$ \\
\hline\hline
\end{tabular}
\end{center}
Notes: $^\sharp$The duration ($\Delta t$) and luminosity ($L_{\rm tot}$) of an FRB and gravitational energy
release rate ($\dot{E}_{\rm G}$) are only expressed as functions of asteroidal mass if typical values of the other
parameters are taken in Table 1, and thus, if the asteroidal mass increases (decreases) by one order of magnitude,
for example, these observed quantities are enhanced (reduced) by factors of 2.78, 7.74 and 3.59, respectively.
$^\P$The indexes of the cumulative duration distribution ($\alpha_T$) and luminosity function ($\alpha_L$) of FRBs are
found to be dependent on the index of the asteroidal differential size distribution ($\beta$) from Equations (2) and (15),
and as in Section 5, it is shown that the observed values of these three indexes for FRB 121102 not only support
our model but also show its self-consistency. $^\|$The repetitive rate (${\cal{R}}_a$) and occurrence rate
(${\cal{R}}_b$) as functions of average proper velocity ($v_*$) and numbers of neutron stars ($N_*$) and asteroid
belts ($N_b$) in a galaxy are only order-of-magnitude estimates, where the repetitive rate is consistent with
the one observed by \cite{Spitler2016} and \cite{Scholz2016} for typical values of all the relevant
parameters in Table 1.
\end{table}


\begin{thebibliography}{99}
\bibitem[Blaes \& Madau(1993)]{Blaes1993} Blaes, O., \& Madau, P. 1993, \apj, 403, 690

\bibitem[Boer et al.(1989)]{Boer1989} Boer, M., Hameury, J. M., \& Lasota, J. P. 1989, \nat, 337, 716

\bibitem[Burke-Spolaor \& Bannister(2014)]{Burke2014} Burke-Spolaor, S., \& Bannister, K.~W. 2014, \apj, 792, 19

\bibitem[Champion et al.(2015)]{Champion2015} Champion, D.~J., Petroff, E., Kramer, M. et al. 2015, \mnras, 460, L30

\bibitem[Chandrasekhar(1960)]{Chandra1960} Chandrasekhar, S. 1960, Plasma Physics (University of Chicago Press, Chicago)

\bibitem[Colgate \& Petscheck(1981)]{Colgate1981} Colgate, S. A., \& Petscheck, A. G. 1981, \apj, 248, 771

\bibitem[Connor et al.(2016)]{Connor2016} Connor, L., Sievers, J., \& Pen, U.-L. 2016, \mnras, 458, L19

\bibitem[Cordes \& Wasserman(2016)]{Cordes2016} Cordes, J.~M., \& Wasserman, I. 2016, \mnras, 457, 232

\bibitem[Davis et al.(2002)]{Davis2002} Davis, D. R., Durda, D. D., Marzari, F., Campo Bagatin, A., \& Gil-Hutton, R. 2002,
Asteroids III (edited by W. F. Bottke Jr., A. Cellino, P. Paolicchi, \& R. P. Binzel, University of Arizona Press, Tucson), p. 545-558

\bibitem[DeMeo \& Carry(2014)]{DeMeo2014} DeMeo, F. E., \& Carry, B. 2014, \nat, 505, 629

\bibitem[Falcke \& Rezzolla(2014)]{Falcke2014} Falcke, H., \& Rezzolla, L. 2014, \aap, 562, A137

\bibitem[Galvez \& Borovsky(1991)]{Galvez1991} Galvez, M., \&  Borovsky, J. E. 1991, Phys. Fluids B, 3, 1892

\bibitem[Geng \& Huang(2015)]{Geng2015} Geng, J.~J., \& Huang, Y.~F. 2015, \apj, 809, 24

\bibitem[Goldreich \& Lynden-Bell(1969)]{GoldreichL1969} Goldreich, P., \& Lynden-Bell, D. 1969, \apj, 156, 59

\bibitem[Goldreich \& Julian(1969)]{GoldreichJ1969} Goldreich, P., \& Julian, W.~H. 1969, \apj, 157, 869

\bibitem[Hansen \& Lyutikov(2001)]{Hansen2001} Hansen, B.~M.~S., \& Lyutikov, M. 2001, \mnras, 322, 695

\bibitem[Harwitt \& Salpeter(1973)]{Harwitt1973} Harwitt, M., \& Salpeter, E. E. 1973, \apj, 186, L37

\bibitem[Huang \& Geng(2014)]{Huang2014} Huang, Y. F., \& Geng, J. J. 2014, \apjl, 782, L20

\bibitem[Ivezi\'c et al.(2001)]{Ivezic2001} Ivezi\'c, Z. et al. 2001, \aj, 122,  2749

\bibitem[Jones(2007)]{Jones2007} Jones, B. W. 2007, Discovering the Solar System (John Wiley \& Sons, Ltd), Chapter 3

\bibitem[Kashiyama et al.(2013)]{Kashiyama2013} Kashiyama, K., Ioka, K., \& M{\'e}sz{\'a}ros, P. 2013, \apjl, 776, L39

\bibitem[Katz(2014)]{Katz2014} Katz, J. I. 2014, \prd, 89, 103009

\bibitem[Katz(2016a)]{Katz2016a} Katz, J. I. 2016a, \apj, accepted, arXiv:1512.04503

\bibitem[Katz(2016b)]{Katz2016b} Katz, J. I. 2016b, Modern Physics Letters A, 31, 1630013

\bibitem[Katz et al.(1994)]{Katz1994} Katz, J. I., Toole, H. A., \& Unruh, S. H. 1994, \apj, 437, 727

\bibitem[Keane et al.(2016)]{Keane2016} Keane, E.~F., Johnston, S., Bhandari, S. et al. 2016, \nat, 530, 453

\bibitem[Keane et al.(2012)]{Keane2012} Keane, E.~F., Stappers, B.~W., Kramer, M., \& Lyne, A.~G. 2012, \mnras, 425, L71

\bibitem[Kulkarni et al.(2014)]{Kulkarni2014} Kulkarni, S.~R., Ofek, E.~O., Neill, J.~D., Zheng, Z., \& Juric, M. 2014, \apj, 797, 70

\bibitem[Lai(2012)]{Lai2012} Lai, D. 2012, \apjl, 757, L3

\bibitem[Litwin \& Rosner(2001)]{Litwin2001} Litwin, C., \& Rosner, R. 2001, \prl, 86, 4745

\bibitem[Liu et al.(2016)]{Liu2016} Liu, T., Romero, G. E., Liu, M. L., \& Li, A. 2016, arXiv:1602.06907

\bibitem[Livio \& Taam(1987)]{Livio1987} Livio, M., \& Taam, R. E. 1987, \nat, 327, 398

\bibitem[Loeb et al.(2014)]{Loeb2014} Loeb, A., Shvartzvald, Y., \& Maoz, D. 2014, \mnras, 439, L46

\bibitem[Lorimer et al.(2007)]{Lorimer2007} Lorimer, D.~R., Bailes, M., McLaughlin, M.~A., Narkevic, D.~J.,
\& Crawford, F. 2007, Science, 318, 777

\bibitem[Lyutikov et al.(2016)]{Lyutikov2016} Lyutikov, M., Burzawa, L., \& Popov, S. B. 2016, arXiv:1603.02891

\bibitem[Madgwick et al.(2002)]{Madgwick2002} Madgwick, D. S., Lahav, O., Baldry, I. K. et al. 2002, \mnras, 333, 133

\bibitem[Masui et al.(2015)]{Masui2015} Masui, K., Lin, H.-H., Sievers, J. et al. 2015, \nat, 528, 523

\bibitem[McWilliams \& Levin(2011)]{McWilliams2011} McWilliams, S.~T., \& Levin, J. 2011, \apj, 742, 90

\bibitem[Mingarelli et al.(2015)]{Mingarelli2015} Mingarelli, C.~M.~F., Levin, J., \& Lazio, T.~J.~W. 2015, \apjl, 814, L20

\bibitem[Mitrofanov \& Sagdeev(1990)]{Mitrofanov1990} Mitrofanov, I. G., \& Sagteev, R. Z. 1990, \nat, 344, 313

\bibitem[Mottez \& Zarka(2014)]{Mottez2014} Mottez, F., \& Zarka, P. 2014, \aap, 569, A86

\bibitem[Neubert et al.(1992)]{Neubert1992} Neubert, T.,  Miller, R. H., Buneman, O., \& Nishikawa, K.-I. 1992, J. Geophys. Res., 97, 12057

\bibitem[Ofek(2009)]{Ofek2009} Ofek, E. O. 2009, Publ. Astron. Soc. Pac., 121, 817

\bibitem[Petroff et al.(2015)]{Petroff2015} Petroff, E., Bailes, M., Barr, E.~D. et al. 2015, \mnras, 447, 246

\bibitem[Piro(2012)]{Piro2012} Piro, A.~L. 2012, \apj, 755, 80

\bibitem[Popov \& Postnov(2010)]{Popov2010} Popov, S.~B., \& Postnov, K.~A. 2010, Evolution of Cosmic Objects
through their Physical Activity (edited by Harutyunian, H. A., Mickaelian, A. M., \& Terzian, Y.), p. 129-132

\bibitem[Punsly \& Bini(2016)]{Punsly2016} Punsly, B., \& Bini, D. 2016, arXiv:1603.05509


\bibitem[Ravi et al.(2015)]{Ravi2015} Ravi, V., Shannon, R.~M., \& Jameson, A. 2015, \apjl, 799, L5

\bibitem[Ruderman \& Sutherland(1975)]{Ruderman1975} Ruderman, M.~A., \& Sutherland, P.~G. 1975, \apj, 196, 51

\bibitem[Safronov(1972)]{Safronov1972} Safronov, V. S. 1972, Evolution of the Protoplanetary Cloud and Formation
of the Earth and Planets (Israel Program for Scientific Translations, Keter Publishing House)

\bibitem[Schmidt(1960)]{Schmidt1960} Schmidt, G. 1960, Phys. Fluids, 3, 961

\bibitem[Scholz et al.(2016)]{Scholz2016} Scholz, P. et al. 2016, arXiv:1603.08880

\bibitem[Shapiro \& Teukolsky(1983)]{Shapiro1983} Shapiro, S.~L., \& Teuklosky, S.~A. 1983, Black Holes, White Dwarfs and Neutron
         Stars: The Physics of Compact Objects (John Wiley \& Sons, New York), p. 241-290

\bibitem[Shull \& Stern(1995)]{Shull1995} Shull, J. M., \& Stern, S. A. 1995, \aj, 109, 690

\bibitem[Spitler et al.(2014)]{Spitler2014} Spitler, L.~G., Cordes, J.~M., Hessels, J.~W.~T. et al. 2014, \apj, 790, 101

\bibitem[Spitler et al.(2016)]{Spitler2016} Spitler, L.~G., Scholz, P., Hessels, J.~W.~T.,
et al. 2016, \nat, 531, 202

\bibitem[Thornton et al.(2013)]{Thornton2013} Thornton, D., Stappers, B., Bailes, M. et al. 2013, Science, 341, 53

\bibitem[Totani(2013)]{Totani2013} Totani, T.\ 2013, \pasj, 65, L12

\bibitem[Van Buren(1981)]{VanBuren1981} Van Buren, D. 1981, \apj, 249, 297

\bibitem[Wang et al.(2016)]{Wang2016} Wang, J. S., Yang, Y. P., Wu, X. F., Dai, Z. G., \& Wang, F. Y. 2016, \apjl, 822, L7

\bibitem[Wang \& Yu(2016)]{WangYu2016} Wang, F. Y., \& Yu, H. 2016, arXiv:1604.08676

\bibitem[Wu et al.(2002)]{Wu2002} Wu, K., Cropper, M., Ramsay, G., \& Sekiguchi, K. 2002, \mnras, 331, 221


\bibitem[Yoshida \& Nakamura(2007)]{Yoshida2007} Yoshida, F., \& Nakamura, T. 2007, Planetary \& Space Science, 55, 1113

\bibitem[Zhang(2014)]{Zhang2014} Zhang, B. 2014, \apjl, 780, L21

\bibitem[Zhang(2016a)]{Zhang2016a} Zhang, B. 2016a, arXiv:1602.04542

\bibitem[Zhang(2016b)]{Zhang2016b} Zhang, B. 2016b, \apjl, 822, L14

\bibitem[Zhou(2016)]{Zhou2016} Zhou, L.-Y. 2016, private communications

\end{thebibliography}
\end{document}